\documentclass[showpacs,aps,amsmath,amssymb]{revtex4}
\usepackage{CJK}
\newcommand{\dbar} {\ensuremath{\,\mathchar'26\mkern-12mu d}}
\usepackage[latin1]{inputenc}
\usepackage{graphicx}
\usepackage{bm}
\usepackage{xcolor}
\usepackage{array}
\usepackage{braket}
\usepackage{mathtools}
\usepackage{microtype}
\usepackage{wasysym}
\renewcommand{\baselinestretch}{1.0}
\begin{document}

\renewcommand{\baselinestretch}{1.0}
\title{Unraveling the role of coherence in the first law of quantum thermodynamics}

\author{Bert\'{u}lio de Lima Bernardo$^{1,2}$}

\affiliation{$^{1}$Departamento de F\'{\i}sica, Universidade Federal da Para\'{\i}ba, 58051-900 Jo\~ao Pessoa, PB, Brazil\\
$^{2}$Departamento de F\'{\i}sica, Universidade Federal de Campina Grande, Caixa Postal 10071, 58109-970 Campina Grande-PB, Brazil}

\email{bertulio.fisica@gmail.com}

\begin{abstract}
                
One of the fundamental questions in the emerging field of quantum thermodynamics is the role played by coherence in energetic processes that occur at the quantum level. Here, we address this issue by investigating two different quantum versions of the first law of thermodynamics, derived from the classical definitions of work and heat. By doing so, we find out that there exists a mathematical inconsistency between both scenarios. We further show that the energetic contribution of the dynamics of coherence is the key ingredient to establish the consistency. Some examples involving two-level atomic systems are discussed in order to illustrate our findings.                                          
   
\end{abstract}

\maketitle


\section{Introduction}

More than a century after the conception of its laws, thermodynamics continues to unveil the underlying nature of many different physical mechanisms \cite{prigogine,landauer,beke,parrondo}. The hallmark of the theory is the effective description of the internal behavior of complex macroscopic systems without taking into consideration the fundamental properties of the microscopic constituents \cite{callen,kittel,reichl}. In general terms, the method through which thermodynamics succeeded in realizing such an impressive task was by focusing on the study of how the total energy of the system can change, and on the constraints imposed by nature about the possible changes. Central to the first question is the first law, which states the conservation of energy: 
``in a system that cannot exchange any matter with the surrounding medium, work and heat are the only two forms of energy transfer'' \cite{kittel}. Work is the transfer of energy that results from changes in the generalized coordinates that characterize the system, such as volume, electric polarization and magnetization \cite{kittel,reichl}. In turn, heat is defined as the energy transfer that accompanies an entropy transfer between the system and its surroundings \cite{kittel,reichl}.    

Besides the astonishing universality of the laws of thermodynamics, another remarkable characteristic of the theory is the demand of so few physical quantities to fully describe a myriad of different processes. Work, heat and temperature are some examples. The reason is that such quantities are a product of the average collective behavior of the system's constituents. Nevertheless, when we deal with systems with a small number of constituents, fluctuations of these thermodynamic quantities become relevant due to the erratic molecular motion, and this is where the field of stochastic thermodynamics comes into play in order to account for these probabilistic aspects \cite{sekimoto,seifert}. In a similar fashion, if we deal with even smaller systems, allied with the thermal fluctuations, quantum effects become prominent, which adds even more unpredictability to the physical quantities involved in the problem. In order to cope with these cases, quantum thermodynamics has emerged with the idea of investigating the laws of thermodynamics in the quantum regime \cite{goold,deffner,vinj,kosloff}, trying to maintain their original simplicity \cite{bert,monsel,alonso,skrz}. To this end, it is crucial to unravel the actual influence of quantum phenomena such as coherence and entanglement on these laws \cite{ali1,aberg,brander,scully,fran,kor,latune,lost,marvi,strel,hilt,santos,beny}.

In this work, we study the extension of the first law of thermodynamics to the quantum domain in two different perspectives: one starting from the classical definition of work and the other from the classical definition of heat. We then observe that the results obtained for the quantum work and heat in their respective scenarios, when put together, are not in agreement with the classical form of the first law. However, we solve this problem by examining the energetic contribution of the dynamics of coherence that occurred in the quantum transformation; an element which is absent in both classical and stochastic thermodynamics. We also demonstrate that the contributions of work, heat and coherence change present a particularly interesting symmetry with respect to the quantum dynamic variables that characterize the process. Our findings are discussed in the light of the normal Zeeman effect, the Rabi oscillation and the spontaneous emission cases, all in the simple framework of two-level atomic systems.

\section{First law from the classical concept of work}

To start with, we survey the main ideas behind the interpretation of the first law of quantum thermodynamics formulated with basis on the classical notion of work, which is largely accepted in the literature \cite{ali2,kieu1,kieu2}.  Let us consider the working substance as an arbitrary quantum system, whose Hamiltonian can be written as $\hat{H} = \sum_{n} E_{n} \ket{n}\bra{n}$, where $E_{n} = \braket{n|\hat{H}|n}$ and $\ket{n}$ are the $n$-th energy eigenvalue and eigenstate, respectively. In this perspective, we can define the internal energy of the system as given by the average of $\hat{H}$, 
\begin{equation}
\label{1}
U = \langle \hat{H} \rangle = tr \{\hat{\rho} \hat{H}  \} = \sum_{n} P_{n} E_{n}, 
\end{equation}
where $\hat{\rho}$ is the density operator of the system, and $P_{n} = \braket{n|\hat{\rho}|n}$ is the probability of the system being in the $n$-th state. In calculating the trace operation in the last equality, the two operators were represented in the energy eigenstate basis $\{ \ket {n} \}$. We also observe that $dU = \sum_{n} [E_{n} d P_{n} + P_{n} d E_{n}]$, from which the work performed and the heat exchange by the working substance in a given infinitesimal transformation can be identified as
\begin{equation}
\label{2}
\dbar W \coloneqq \sum_{n} P_{n} d E_{n} 
\end{equation}
and 
\begin{equation}
\label{3}
\dbar Q \coloneqq \sum_{n} E_{n} d P_{n}, 
\end{equation}
respectively \cite{ali2,kieu1,kieu2,quan}. 

In order to make such an intuitive identification, we implicitly invoke the classical concept of work: ``the work realized on or by the working substance is the change in the internal energy produced by modifications in the generalized coordinates'' \cite{kittel,reichl,landau}. In terms of quantum mechanics, such a modification in the generalized coordinates naturally causes alterations in the energy level configuration $E_{n}$, which justifies Eq.~(\ref{2}). 
Having established this point, we are left with the definition of heat, according to Eq.~(\ref{3}), as being a result of variations in the occupation probabilities of the energy levels, $P_{n}$. This is a reasonable interpretation, but it is not a direct extension of the classical concept of heat, as we shall see. In any case, taken together, these two relations provide a quantum version of the first law analogous to that of classical thermodynamics, $dU = \dbar W + \dbar Q$. 

To illustrate these definitions, let us consider the important case of an isothermal process of a quantum system in thermal equilibrium with a heat bath at a temperature $T$ \cite{quan,ribeiro}. In this case, we assume the initial state of the system as the (thermal) Gibbs state \begin{equation}
\label{4}
\hat{\rho}_{th} = \sum_{n} \frac{e^{-\beta E_{n}}}{Z} \ket{n} \bra{n}, 
\end{equation}
where $Z = \sum_{n} e^{-\beta E_{n}}$ is the partition function and $\beta = 1/ k_{B} T$, with $k_{B}$ the Boltzmann's constant. We also note that $P_{n} = e^{-\beta E_{n}}/Z$. For this equilibrium state we can also define the free energy as $F = - k_{B} T ln (Z)$, which provides $dF = -k_{B}T dZ/Z = \sum_{n} P_{n} dE_{n}$. By comparison of this result and Eq.~(\ref{2}) we can write that
\begin{equation}
\label{5}
\dbar W = dF, 
\end{equation}
as expected for an infinitesimal isothermal process. On the other hand, in what concerns the above definition of heat, we observe that $E_{n} = -k_{B} T ln (Z P_{n})$, which, if substituted into Eq.~(\ref{3}), yields $\dbar Q = -k_{B} T \sum_{n} ln (P_{n}) dP_{n}$, where we used the fact that $\sum_{n} d P_{n} = 0$, because $\sum_{n} P_{n} = 1$. For this case, in which the density operator is diagonal in the energy eigenstate basis, the von Neumann entropy of the system can be written simply as
\begin{equation}
\label{6}
S = -k_{B} \sum_{n} P_{n} ln (P_{n}). 
\end{equation}
This allows us to write $dS = -k_{B} \sum_{n} ln (P_{n}) dP_{n}$, where we used again the fact that $\sum_{n} d P_{n} = 0$. Therefore, we find that
\begin{equation}
\label{7}
\dbar Q = T dS, 
\end{equation}
which is also the classical result for an infinitesimal isothermal transformation. In the end, Eqs.~(\ref{5}) and~(\ref{7}) provided results for a quantum isothermal process which are consistent with the definitions of work and heat of the classical counterpart.

\section{First law from the classical concept of heat}

Here we want to revisit the formulation of the first law in the quantum realm, but now based on a extension of the classical notion of heat. Recently, some works have addressed this issue based on a similar perspective \cite{alipour,ahmadi}. To begin with, we reevaluate the internal energy of the working substance as in Eq.~(\ref{1}). However, instead of using the energy eigenstate basis $\{ \ket {n} \}$ to calculate the trace, we use the eigenstate basis of the density operator $\{ \ket {k} \}$, which in general is different from $\{ \ket{n} \}$,  
\begin{equation}
\label{8}
U= \langle \hat{H} \rangle = tr \{\hat{\rho} \hat{H}  \} = \sum_{k} \rho_{k} \epsilon_{k}, 
\end{equation}
with $\rho_{k} = \braket{k|\hat{\rho}|k}$ being the eigenvalues of $\hat{\rho}$, i.e., $\hat{\rho} = \sum_{k} \rho_{k} \ket{k}\bra{k}$, and $\epsilon_{k} = \braket{k|\hat{H}|k}$ the diagonal elements of $\hat{H}$ represented in the $\{ \ket{k} \}$ basis. From Eq.~(\ref{8}), we have that 
\begin{equation}
\label{11}
dU = \sum_{k} [\epsilon_{k} d\rho_{k} + \rho_{k} d\epsilon_{k}]. 
\end{equation}
We also have that the von Neumann entropy of the system is given by $S=-k_{B} tr \{ \hat{\rho} log \hat{\rho}\} = - k_{B} \sum_{k} \rho_{k} log \rho_{k}$. Accordingly, we obtain that
\begin{equation}
\label{12}
dS = - k_{B}\sum_{k} [log (\rho_{k}) d\rho_{k}], 
\end{equation}
where we used the fact that $\sum_{k} d \rho_{k} = 0$, once  $\sum_{k} \rho_{k} = 1$, which holds whenever the system evolves under a trace-preserving quantum operation.

At this point, we invoke the classical concept of heat: ``the heat exchanged between the working substance and the external environment corresponds to the change in the internal energy that is accompanied by entropy change'' \cite{kittel,landau}. In terms of quantum mechanics, this definition together with Eq.~(\ref{12}) tell us that the contribution of heat to the change in the internal energy of the system is revealed to exist only if $d \rho_{k} \neq 0$. Therefore, the extension of the classical concept of heat leads us to identify work and heat in the quantum domain alternatively as 
\begin{equation}
\label{13}
\dbar \mathcal{W} \coloneqq \sum_{k} \rho_{k} d \epsilon_{k} 
\end{equation}
and 
\begin{equation}
\label{14}
\dbar \mathcal{Q} \coloneqq \sum_{k} \epsilon_{k} d \rho_{k}, 
\end{equation}
respectively. These definitions also allow us to write a quantum version of the first law, $dU = \dbar \mathcal{W} + \dbar \mathcal{Q}$. 

As an application of this alternative version, let us once again study the case of an isothermal quantum process. The analysis becomes trivial if we observe that  the system is described by a thermal state as that of Eq.~(\ref{4}) during the entire transformation, and that this state is diagonal both in the $\{ \ket{n} \}$ and $\{ \ket{k} \}$ bases. In this specific case, it is easy to see that $E_{n} = \epsilon_{k}$ and $P_{n} = \rho_{k}$, so that $\dbar \mathcal{W} = \dbar W$ and $\dbar \mathcal{Q} = \dbar Q$. That is, the definitions of work and heat are the same in both perspectives. As a result, Eqs.~(\ref{5}) and~(\ref{7}) are also recovered in this alternative formulation of the first law. 

\section{The role of coherence}

In classical thermodynamics, the concepts of work and heat, as presented in the introduction, are consistent with each other in unambiguously separating on phenomenological grounds the two possible contributions to a change in the internal energy of the system. We also have seen above that this consistency encompasses the case of quantum isothermal processes. In this context, we now pose the question of whether this is the case that these classical concepts of work and heat can always be promptly extended to the quantum domain, while keeping the structure of the first law. In order to answer this question, we need to investigate the equivalence between the two mathematical formulations of work and heat derived above for a general quantum process.     

Let us first investigate whether there is some correspondence between the two work expressions discussed above. The one obtained from the classical concept of work, Eq.~(\ref{2}), can be rewritten as 
\begin{align}
\label{17}
\dbar W & = \sum_{n} P_{n} d E_{n} = \sum_{n} \braket{n|\left(\sum_{k} \rho_{k} \ket{k}\bra{k}\right)|n} d E_{n}  \nonumber \\
&= \sum_{n} \sum_{k} \rho_{k} |c_{n,k}|^{2} d E_{n},
\end{align}
with $c_{n,k} = \braket{n|k}$. On the other hand, the work expression derived based on the classical concept of heat, Eq.~(\ref{13}), yields
\begin{align}
\label{18}
\dbar \mathcal{W} & = \sum_{k} \rho_{k} d \epsilon_{k} = \sum_{k} \rho_{k}  d \left[ \braket{k| \left(\sum_{n} E_{n} \ket{n} \bra{n} \right)|k} \right]   \nonumber \\
&= \sum_{k} \rho_{k}  d \left[  \sum_{n} E_{n} |c_{n,k}|^{2} \right] \nonumber \\
&= \sum_{n} \sum_{k} \rho_{k} |c_{n,k}|^{2} d E_{n} + \sum_{n} \sum_{k} (E_{n} \rho_{k}) d \left[ |c_{n,k}|^{2} \right] \nonumber \\
&= \dbar W + \dbar \mathcal{C},
\end{align}
where in the last equality we used the result of Eq.~(\ref{17}), and defined 
\begin{equation}
\label{19}
\dbar \mathcal{C} = \sum_{n}\sum_{k} (E_{n} \rho_{k}) d \left[ |c_{n,k}|^{2} \right].
\end{equation}
This path-dependent contribution to the change of the internal energy represents the quantitative difference between the two expressions of the quantum work obtained from the classical concepts of work and heat. 

Before commenting further on the quantity of Eq.~(\ref{19}), let us move on to the comparison between the two expressions of quantum heat discussed here. The one obtained with basis on the classical concept of heat, Eq.~(\ref{14}), can be rewritten as
\begin{align}
\label{20}
\dbar \mathcal{Q} & = \sum_{k} \epsilon_{k} d \rho_{k} = \sum_{k} \braket{k|\left(\sum_{n} E_{n} \ket{n}\bra{n}\right)|k} d \rho_{k}  \nonumber \\
&= \sum_{n} \sum_{k} E_{n} |c_{n,k}|^{2} d \rho_{k}.
\end{align}
Conversely, the other obtained from the classical concept of work, Eq.~(\ref{3}), provides that
\begin{align}
\label{21}
\dbar Q & = \sum_{n} E_{n} d P_{n} = \sum_{n} E_{n}  d \left[ \braket{n| \left(\sum_{k} \rho_{k} \ket{k} \bra{k} \right)|n} \right]   \nonumber \\
&= \sum_{n} E_{n}  d \left[  \sum_{k} \rho_{k} |c_{n,k}|^{2} \right] \nonumber \\
&= \sum_{n}\sum_{k} E_{n} |c_{n,k}|^{2} d \rho_{k} + \sum_{n}\sum_{k} (E_{n} \rho_{k}) d \left[ |c_{n,k}|^{2} \right] \nonumber \\
&= \dbar \mathcal{Q} + \dbar \mathcal{C},
\end{align}
where in the last equality we used the definitions of Eqs.~(\ref{19}) and~(\ref{20}).

Having completed our comparative analysis, it is now clear from Eqs.~(\ref{18}) and~(\ref{21}) that the two quantum extensions of the first law of thermodynamics are not equivalent. Furthermore, the difference lies essentially in how the contribution of $\dbar \mathcal{C}$ to the change of the internal energy of the system is categorized; whether as work or heat. The formalism based on the classical concept of work considers $\dbar \mathcal{C}$ strictly as heat, Eq.~(\ref{21}), whereas the formalism based on the classical concept of heat considers it as work, Eq.~(\ref{18}). As can be seen, it is essential to examine the physical origin of $\dbar \mathcal{C}$ to solve this conundrum. From Eq.~(\ref{19}) we see that $\dbar \mathcal {C}$ does not depend on either $d E_{n}$ or $d \rho_{k}$, which would lead us to directly classify it as work or heat, respectively. Instead, it depends on the variation of the quantity $|c_{n,k}(t)|
^2 = |\braket{n(t)|k(t)}|
^2$. In a quantum process, this quantity varies only if the directions of the basis vectors $\ket{k}$ of the density operator change with respect to the basis vectors $\ket{n}$ of the Hamiltonian. Physically, such a variation occurs if the {\it quantum coherence} of the system (in the energy eigenstate basis) changes with time \cite{strel,baum}. 

In this form, we see that the dynamics of coherence plays a fundamental and exclusive role in the quantum version of the first law of thermodynamics, which is independent of those of the quantum work and heat derived from their respective classical analogues. The energetic contribution of the dynamics of coherence with time in a finite quantum process is obtained by integration of Eq.~(\ref{19}),
\begin{align}
\label{22}
\mathcal{C}(t)
&= \sum_{n}\sum_{k} \int_{0}^{t} (E_{n} \rho_{k}) \frac{d}{dt'} |c_{n,k}|^{2} dt'.
\end{align}
This expression can be in principle calculated if we are given $\hat{\rho}(t)$ and $\hat{H}(t)$. Having found out the role of coherence in the first law, we are left with the definitions of quantum work and heat, which are obtained directly from their original classical concepts, Eqs.~(\ref{17}) and~(\ref{20}). The time-dependence of these quantities in a finite quantum process can also be calculated by direct integration:
\begin{equation}
\label{22.1}
W(t)  =  \sum_{n} \sum_{k} \int_{0}^{t}  \rho_{k} |c_{n,k}|^{2} \frac{d E_{n}}{dt'} dt',
\end{equation}
\begin{equation}
\label{22.2}
\mathcal{Q}(t)  =  \sum_{n} \sum_{k} \int_{0}^{t}  E_{n} |c_{n,k}|^{2} \frac{d \rho_{k}}{dt'} dt'.
\end{equation}
As can be seen, there exists a remarkable symmetry in Eqs.~(\ref{22}),~(\ref{22.1}) and~(\ref{22.2}) with respect to the dependence on the quantum dynamic elements $E_{n}(t)$, $\rho_{k}(t)$ and $|c_{n,k}(t)|^2$.
In parallel with $W$ and $\mathcal{Q}$, the path-dependent quantity $\mathcal{C}$ unambiguously represents the role played by coherence in the first law. In addition, since $\mathcal{C}$ does not have a classical analogue such as $W$ and $\mathcal{Q}$, here we propose a redefinition of the first law of quantum thermodynamics as  
\begin{equation}
\label{23}
d U = \dbar W + \dbar \mathcal{Q} + \dbar \mathcal{C}.
\end{equation}
Note also that the change in the internal energy acquires an interesting form,
\begin{equation}
\label{22.3}
\Delta U(t)  =  \sum_{n} \sum_{k} \int_{0}^{t} \frac{d}{dt'} \left(E_{n} \rho_{k} |c_{n,k}|^{2} \right)  dt'.
\end{equation}
In Eq.~(\ref{23}), we separate the semiclassical contributions of work and heat to $dU$ from the purely quantum mechanical contribution due to the dynamics of coherence. This is our main result, which shall be examined on the basis of some examples. 

\section{Examples}

We now apply our findings to three well-known processes that occur in atomic systems. In the following examples, the working substance is considered to be a two-level atom, whose ground and excited states, $\ket{g}$ and $\ket{e}$, have energies $E_{g}$ and $E_{e}$, respectively, so that the Hamiltonian is given by $\hat{H}_{S} = E_{g}\ket{g}\bra{g} + E_{e}\ket{e}\bra{e}$. The first process to be studied is the {\it normal Zeeman effect}. In this case, if we suppose that the atom is in the excited state and that the magnetic quantum numbers of $\ket{g}$ and $\ket{e}$ are respectively $m = 0$ and $m = 1$, the application of an external magnetic field of magnitude $B$ causes a shift in the transition energy of $\Delta U = (e \hbar/ 2 m_{e}) B$, where $e$ and $m_{e}$ are the charge and mass of the electron, respectively, and $\hbar$ is Planck's constant \cite{foot,cohen}. From our viewpoint, the application of the magnetic field causes a change in the energy level configuration ($d E_{n} \neq 0$), which can be interpreted as realization of work on the atom. Still, due to the absence of entropy and coherence changes ($d \rho_{k}= d  |c_{n,k}|^{2} = 0$), we have that $\Delta U = W = (e \hbar/ 2 m_{e}) B$ and $\mathcal{Q} = \mathcal{C} = 0$. 

Our second example is the {\it Rabi oscillation}. This process takes place when the atom is in the presence of a strong resonant electromagnetic field. In this case, if we consider that the atom starts out in the ground state at $t=0$, its time evolution is given by $\ket{\psi(t)} = \cos(\Omega_{R}t/2) \ket{g} + i \sin(\Omega_{R}t/2) \ket{e}$, where $\Omega_{R}$ is the so-called Rabi frequency \cite{foot,cohen,fox}. This corresponds to a pure state evolution, $\hat{\rho}(t) = \ket{\psi(t)}\bra{\psi(t)}$, that together with the Hamiltonian of the system $\hat{H}_{S}$ allows us to study the first law. Since the energy levels are fixed and the state is pure along the entire quantum dynamics ($d E_{n} = d \rho_{k}$ = 0), we have that $W = \mathcal{Q} = 0$. In turn, by using Eq.~(\ref{22}) we find that $\mathcal{C}(t) = [\cos
^{2}(\Omega_{R}t/2)-1]E_{g} + [\sin
^{2}(\Omega_{R}t/2)]E_{e}$. Thus, the only contribution to the change in the internal energy is due to the dynamics of coherence, $\Delta U (t)= \mathcal{C}(t)$ \cite{SM}. Fig. 1 illustrates this behavior. 
\begin{figure}[ht]
\centerline{\includegraphics[width=9.2cm]{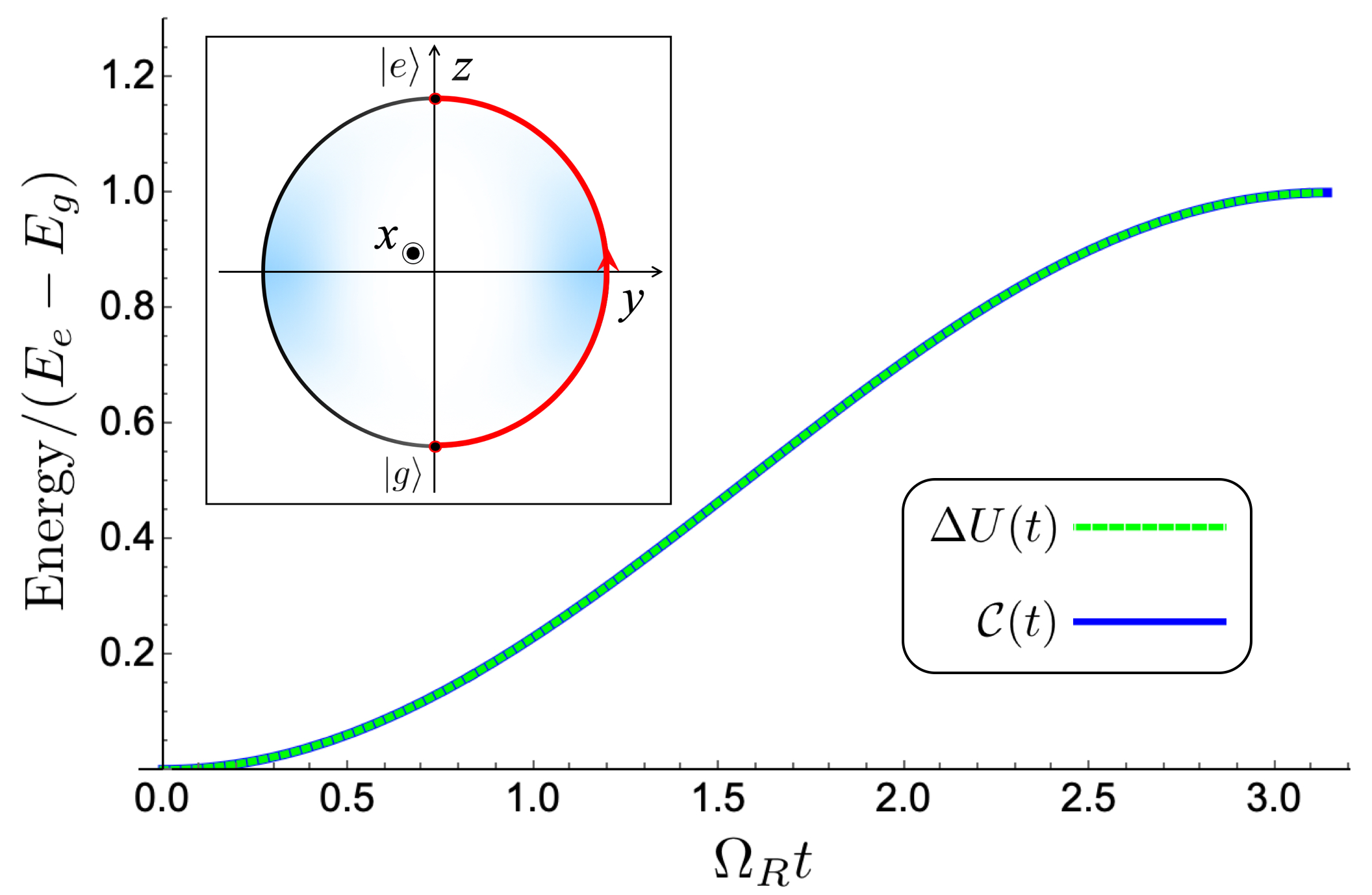}}
\caption{ (color online) First law description of the Rabi oscillation, in which an external field causes an unitary transformation in the state of the atom from $\ket{g}$ to $\ket{e}$. The change in the internal energy is completely related to the dynamics of coherence, $\Delta U (t)= \mathcal{C} (t)$. Inset: Bloch sphere representation of the process in the $yz$ plane. The blue regions indicate higher coherence.}
\label{setup}
\end{figure}

As a last example, we consider the problem of the {\it spontaneous emission} of a photon \cite{cohen, fox}, in which case the atom is assumed to be initially prepared in the state $\ket{\psi(0)} = 1/\sqrt{2}(\ket{g}+\ket{e})$, and the time evolution described by means of the amplitude-damping channel \cite{nielsen,preskill}. In this case, if we suppose that the probability of a decaying event per unit time is $\Gamma$, the density operator as a function of time, in the energy basis $\{\ket{g}, \ket{e}\}$, is given by \cite{SM}
\begin{equation}
\label{25}
\hat{\rho}(t) = \frac{1}{2}
\begin{pmatrix}
2 - e^{-\Gamma t} & e^{-\Gamma t/2} 
 \\
e^{-\Gamma t/2} & e^{-\Gamma t}  
\end{pmatrix}.
\end{equation}
Again, with this density matrix along with the Hamiltonian of the system, $\hat{H}_{S}$, describing the quantum dynamics, all relevant functions of Eq.~(\ref{23}) can be evaluated for this process. Since the energy eigenvalues are constant ($d E_{n} = 0$), obviously no work is done, $W = 0$. Nevertheless, if we calculate the eigenvalues and eigenstates of $\hat{\rho}(t)$, we can obtain $\mathcal{C}(t)$ and $\mathcal{Q}(t)$ by means of Eqs.~(\ref{22}) and~(\ref{22.2}) \cite{SM}. The results presented in Fig. 2 show that heat is first released and then absorbed by the atom, which is a reflection of the entropy oscillation \cite{comment}. The process also causes a maximum extraction of coherence from the atom, which renders a prominent contribution to the decrease in the internal energy.     
\begin{figure}[ht]
\centerline{\includegraphics[width=9.4cm]{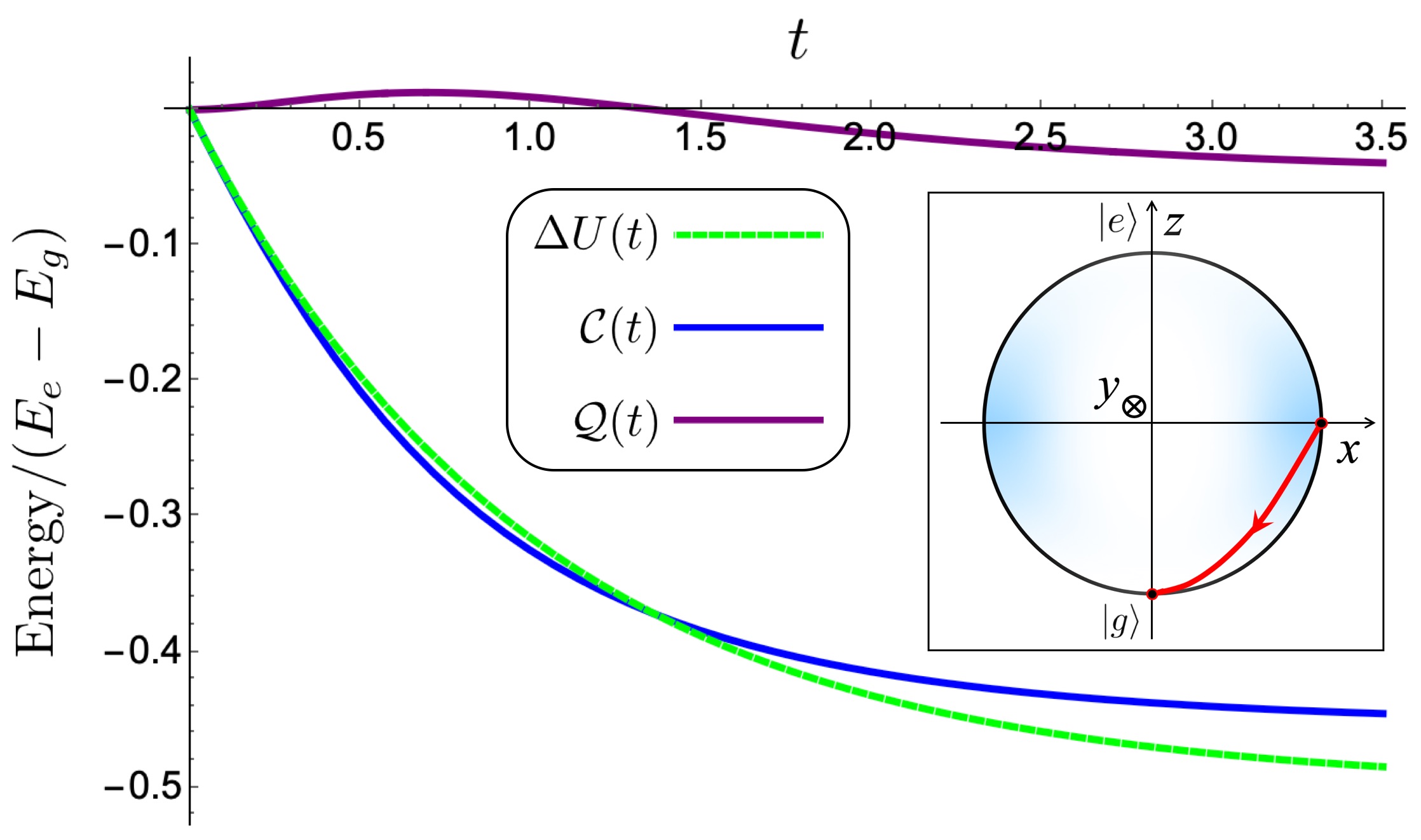}}
\caption{ (color online) First law description of the non-unitary spontaneous emission process. In the quantum dynamics from $\ket{\psi(0)} = 1/\sqrt{2} (\ket{g}+\ket{e})$ to $\ket{g}$, both heat and the dynamics of coherence contribute to the change in the internal energy, $\Delta U (t) = \mathcal{Q}(t) + \mathcal{C}(t)$. We assumed $\Gamma=1$ for simplicity. Inset: Bloch sphere representation of the process in the $xz$ plane. The blue regions indicate higher coherence.}
\label{setup}
\end{figure}

\section{Outlook and Conclusions}

Let us now discuss our approach to the first law of quantum thermodynamics in connection with others developed in recent years. We must call attention to the fact that all energetic contributions proposed here, Eqs. (\ref{22}) to (\ref{22.2}), were derived from the assumption that the internal energy of a given quantum system is equivalent to the average of its energy, which is an idea originally proposed by Alicki \cite{ali2}. We then determined what portion of the internal energy could be considered as quantum work and heat, based on the definitions of their classical counterparts, from which we could explicitly work out the role played by coherence in the first law. Given this line of reasoning, we must also interpret Eqs. (\ref{22}) to (\ref{22.2}) as the average contributions of coherence, work and heat to the change in the internal energy of the quantum system, irrespective of whether the transformation protocol is an equilibrium or nonequilibrium process. Of course, these averages are quantities that become important in the so-called {\it many-runs} regime \cite{kor}, i.e., when a large number of realizations of the transformation protocol is performed on an ensemble of equally prepared {\it individual} systems.

The study of this regime is particularly useful in describing quantum heat engines \cite{quan,kosloff2}. Moreover, this is the scenario in which are applicable the fluctuation relations, such as Jarzynski's equality \cite{jarz,talkner} and Crooks' theorem \cite{crooks}, which relate nonequilibrium fluctuating quantities with equilibrium state functions \cite{campisi,beyer}. In this context, some relevant questions arise, e.g., how the present results could be extended to the single-shot regime, in which the fluctuations are very large, and a probabilistic treatment is required. Such a study would also allow making a direct comparison with the resource theory approach to thermodynamics \cite{lostaglio}.

It is also important to point out that the thermodynamic quantities studied here were conveniently derived by investigating the system, without considering directly either the characteristics of the environment or the system-environment interaction. However, interesting thermodynamic and information-theoretic effects appear when such elements are taken into account \cite{bera}. For instance, in the strong coupling regime of interaction, in which entanglement is established between the system and environment, the amount of heat leaving the system may be very different from that which enters the environment. In fact, these quantities are equivalent only in the weak coupling regime, when system and environment can be considered as uncorrelated during the entire transformation process. These effects can be better understood if we recall the subadditivity of the von Neumann entropy for composite quantum systems \cite{nielsen}.

In conclusion, we have used the very notions of work and heat of classical thermodynamics, which have their origins respectively in the changes of the generalized coordinates and entropy of the system, to propose analogous counterparts in the quantum regime. Having these definitions and investigating their action on the first law, we demonstrate that the role played by quantum coherence in the change of the internal energy has an origin independent of those of work and heat. Evidently, this physical influence due to coherence has no place in classical processes. Further, we quantitatively demonstrated the contributions of work, heat and coherence dynamics in the first law, and used them to study some quantum transformations in the framework of a two-level atom. The present approach sheds a new light on the issue of harnessing coherence for applications in future technologies based on quantum thermodynamic systems.

\section*{Acknowledgements}

The author acknowledges financial support from the
Brazilian funding agencies Coordena{\c c}{\~a}o de Aperfei{\c c}oamento
de Pessoal de N{\'i}vel Superior (CAPES, Finance Code 001), Conselho Nacional de Desenvolvimento Cient{\'i}fico e Tecnol{\'o}gico (CNPq, Grant No. 303451/2019-0), and Pronex/Fapesq-PB/CNPq (Grant Number 0016/2019).

\appendix

\section{Thermodynamics of the Rabi Oscillation}

In this section, we detail the study of the quantum version of the first law of thermodynamics for the Rabi oscillation case, according to Eq.(\ref{23}). As pointed out in the main text, the time evolution of the atom, initially prepared in the state $\ket{g}$, is given by 
\begin{equation}
\label{S1}
\ket{\psi(t)} = \cos(\Omega_{R}t/2) \ket{g} + i \sin(\Omega_{R}t/2) \ket{e},
\end{equation}
which provides the density operator
\begin{equation}
\label{S2}
\hat{\rho}(t) = \cos^{2}(\Omega_{R}t/2) \ket{g}\bra{g} +  \sin^2(\Omega_{R}t/2) \ket{e}\bra{e} - \frac{i}{2}\sin(\Omega_{R}t) \ket{g}\bra{e} +\frac{i}{2}\sin(\Omega_{R}t) \ket{e}\bra{g}.
\end{equation}
The Hamiltonian of the system is
\begin{equation}
\label{S3}
\hat{H}_{S} = E_{g}\ket{g}\bra{g} + E_{e}\ket{e}\bra{e}.
\end{equation}
From Eq.~(\ref{S2}), the eigenvectors of $\hat{\rho}(t)$ can be found to be
\begin{equation}
\label{S4}
\ket{k_{0}(t)} = \cos(\Omega_{R}t/2) \ket{g} + i \sin(\Omega_{R}t/2) \ket{e}
\end{equation}
and
\begin{equation}
\label{S5}
\ket{k_{1}(t)} = \sin(\Omega_{R}t/2) \ket{g} - i \cos(\Omega_{R}t/2) \ket{e},
\end{equation}
with the respective time-independent eigenvalues $\rho_{0} = 1$ and $\rho_{1} = 0$. It is easy to see that the energy eigenvalues are $E_{g}$ and $E_{e}$.

At this point, we are now in a position to calculate the energetic contribution due to the dynamics of coherence in this process, according to Eq.~(\ref{22}):
\begin{align}
\label{S6}
\mathcal{C}(t)
&= \sum_{n}\sum_{k} \int_{0}^{t} (E_{n} \rho_{k}) \frac{d}{dt'} |c_{n,k}|^{2} dt' \nonumber \\
&= E_{g} \int_{0}^{t}  \frac{d}{dt'} | \braket{g|k_{0}(t')} |^{2} dt' + E_{e} \int_{0}^{t} \frac{d}{dt'} | \braket{e|k_{0}(t')}|^{2} dt'\nonumber \\
&= E_{g} \int_{0}^{t} \frac{d}{dt'} | \cos(\Omega_{R}t'/2) |^{2} dt' + E_{e} \int_{0}^{t} \frac{d}{dt'} | i \sin(\Omega_{R}t'/2) |^{2} dt'\nonumber \\
& = E_{g} [\cos
^{2}(\Omega_{R}t/2)-1] + E_{e} [\sin
^{2}(\Omega_{R}t/2)].
\end{align}
This is the result presented in the main text. The other results, $W = \mathcal{Q} = 0$ and $\Delta U (t)= \mathcal{C}(t)$, are clearly seen from Eqs.~(\ref{22.1}) to (\ref{22.3}), and the fact that $d E_{n} = d \rho_{k} = 0$.

\section{Thermodynamics of the Spontaneous emission}

\subsection{Quantum Dynamics}

Now we detail the first law analysis for the spontaneous emission case. First, we derive the time-dependent density operator, which is obtained by means of the study of the amplitude-damping channel. The dynamics of the system is described by the relations \cite{preskill}
\begin{equation}
\label{S7}
\ket{g,E_{0}} \rightarrow \ket{g,E_{0}},  
\end{equation}
\begin{equation}
\label{S8}
\ket{e,E_{0}} \rightarrow \sqrt{1-p} \ket{e,E_{0}} + \sqrt{p} \ket{g,E_{1}}.  
\end{equation}
Eq.~(\ref{S7}) indicates that if the atom is initially in the ground state $\ket{g}$, and the environment is in a given initial state $\ket{E_{0}}$, the joint state $\ket{g,E_{0}}$ does not evolve. On the contrary, if the atom starts out in the excited state $\ket{e}$ with the environment in the initial state $\ket{E_{0}}$, which yields the joint state $\ket{e,E_{0}}$, there exists a probability $p$ that, after a given amount of time, the atom decays to the ground state emitting a photon to the environment, $\ket{g,E_{1}}$, and a probability $1-p$ that the joint system remains unchanged, $\ket{e,E_{0}}$.    

Taken together, Eqs.~(\ref{S7}) and (\ref{S8}) allow us to write the unitary transformation operator 
\begin{equation}
\label{S9}
\hat{U} = \ket{g,E_{0}} \bra{g,E_{0}} + \sqrt{1-p} \ket{e,E_{0}} \bra{e,E_{0}} + \sqrt{p} \ket{g,E_{1}} \bra{e,E_{0}}.
\end{equation}
In turn, the Kraus operators $\hat{K}_{i} = \braket{E_{i}|\hat{U}|E_{0}}$ are given by $\hat{K}_{0} = \braket{E_{0}|\hat{U}|E_{0}} = \ket{g}\bra{g} + \sqrt{1-p} \ket{e}\bra{e}$ and $\hat{K}_{1} = \braket{E_{1}|\hat{U}|E_{0}} =  \sqrt{p} \ket{g}\bra{e}$. Therefore, we can now write the evolution of the density operator in the Kraus representation as
\begin{equation}
\label{S10}
\hat{\rho} (0) \rightarrow \mathcal{E} [\hat{\rho}(0)] = \sum_{i} \hat{K}_{i} \hat{\rho} (0) \hat{K}^{\dagger}_{i},
\end{equation}
or
\begin{equation}
\label{S11}
\hat{\rho}(0) = 
\begin{pmatrix}
\rho_{00} & \rho_{01}
 \\
\rho_{10} & \rho_{11}  
\end{pmatrix}
\rightarrow
\mathcal{E} [\hat{\rho}(0)] =
\begin{pmatrix}
\rho_{00} + p \rho_{11} & \sqrt{1-p} \rho_{01}
 \\
\sqrt{1-p} \rho_{10} & (1-p)\rho_{11}  
\end{pmatrix}.
\end{equation}

We now consider that the probability of a decaying event per unit time is $\Gamma$, so that $p = \Gamma \Delta t \ll 1$ for a time interval $\Delta t$, and the evolution after a time $t = n \Delta t$ is a result of the operation $\mathcal{E}^n [\hat{\rho}(0)]$. In this case, the probabilistic factors in Eq.~(\ref{S11}) are transformed according to $(1-p) \rightarrow (1-p)^{n} = \lim_{n \to \infty} \left(1-\frac{\Gamma t}{n} \right)^n = e
^{- \Gamma t}$, where we assumed $\Delta t \rightarrow 0$. In doing so, we have that the time evolution of the density operator is given by
\begin{equation}
\label{S12}
\hat{\rho}(t) =
\begin{pmatrix}
\rho_{00} + (1-e^{-\Gamma t}) \rho_{11} & e^{-\Gamma t/2} \rho_{01}
 \\
e^{-\Gamma t/2} \rho_{10} & e^{-\Gamma t} \rho_{11} 
\end{pmatrix}.
\end{equation}
Finally, since the initial state is $\ket{\psi(0)} = 1/\sqrt{2} (\ket{g}+\ket{e})$, which provides $\rho_{00}=\rho_{01}=\rho_{10}=\rho_{11}=1/2$, we have that
\begin{equation}
\label{S13}
\hat{\rho}(t) = \frac{1}{2}
\begin{pmatrix}
2 - e^{-\Gamma t} & e^{-\Gamma t/2} 
 \\
e^{-\Gamma t/2} & e^{-\Gamma t}  
\end{pmatrix}.
\end{equation}
This is the time-dependent density operator describing the dynamics of the spontaneous emission in Eq.~(\ref{25}).

\subsection{First Law Description}

Besides the time-dependent density operator of Eq.~(\ref{S13}), we also have the two-level Hamiltonian $\hat{H}_{S} = E_{g}\ket{g}\bra{g} + E_{e}\ket{e}\bra{e}$ in this case. Since the energy eigenvalues are not time-dependent, it is straightforward from Eq.~(\ref{22.1}) that $W = 0$. However, in order to evaluate $\mathcal{Q}(t)$ and $\mathcal{C}(t)$, we must calculate the eigenvalues of $\hat{\rho}(t)$, which can be found to be
\begin{equation}
\label{S14}
\rho_{0}(t) = \frac{1}{2} e^{- \Gamma t} \left(e^{\Gamma t}+\sqrt{-e^{ \Gamma t} +e^{2 \Gamma t}+1}\right)
\end{equation}
and
\begin{equation}
\label{S15}
\rho_{1}(t) = \frac{1}{2} e^{- \Gamma t} \left(e^{\Gamma t}-\sqrt{-e^{ \Gamma t} +e^{2 \Gamma t}+1}\right),
\end{equation}
as well as the respective eigenvectors\begin{equation}
\label{S16}
\ket{k_{0}(t)} =  \frac{\left[e^{-\Gamma t/2} \left(\sqrt{-e^{\Gamma t}+e^{2 \Gamma t}+1}+e^{\Gamma t}-1\right)\ket{g} + \ket{e} \right] }{\sqrt{e^{- \Gamma t} \left(\sqrt{-e^{\Gamma t}+e^{2 \Gamma t}+1}+e^{\Gamma t}-1\right)^2+1}}
\end{equation}
and
\begin{equation}
\label{S17}
\ket{k_{1}(t)} = \frac{\left[ e^{- \Gamma t/2} \left(-\sqrt{-e^{\Gamma t} +e^{2 \Gamma t}+1}+e^{\Gamma t}-1\right)\ket{g} + \ket{e} \right]}{\sqrt{e^{- \Gamma t} \left(-\sqrt{-e^{\Gamma t}+e^{2 \Gamma t}+1}+e^{\Gamma t}-1\right)^2+1}}.
\end{equation}

With the results of Eqs.~(\ref{S14}) to~(\ref{S17}), we can calculate the heat exchanged between the atom and the environment as a function of time by means of Eq.~(\ref{22.2}),
\begin{align}
\label{S18}
\mathcal{Q}(t)  &=  \sum_{n} \sum_{k} \int_{0}^{t}  E_{n} |c_{n,k}|^{2} \frac{d \rho_{k}}{dt'} dt' \nonumber \\
&= E_{g} \left[ \int_{0}^{t}  |\braket{g|k_{0}(t')}|^{2} \frac{d}{dt'} \rho_{0} (t') dt' + \int_{0}^{t}  |\braket{g|k_{1}(t')}|^{2} \frac{d}{dt'} \rho_{1} (t') dt' \right]\nonumber \\ &+E_{e} \left[ \int_{0}^{t}  |\braket{e|k_{0}(t')}|^{2} \frac{d}{dt'} \rho_{0} (t') dt' + \int_{0}^{t}  |\braket{e|k_{1}(t')}|^{2} \frac{d}{dt'} \rho_{1} (t') dt' \right] \nonumber \\
&= \frac{1}{4} (E_{e}-E_{g}) \left[2 e^{-t}-\frac{1}{2} \log \left(e^{-2 t}-e^{-t}+1\right)-\sqrt{3} \tan
   ^{-1}\left(\frac{2 e^{-t}-1}{\sqrt{3}}\right)+\frac{\pi }{2 \sqrt{3}}-2 \right],
\end{align}
where we have assumed $\Gamma =1$ for simplicity. This result is plotted in Fig. 2.

In what follows, we calculate the energetic contribution of the dynamics of coherence in this example. By assuming $\Gamma =1$ again, from Eq.~(\ref{22}) we obtain that 
\begin{align}
\label{S19}
\mathcal{C}(t)
&= \sum_{n}\sum_{k} \int_{0}^{t} (E_{n} \rho_{k}) \frac{d}{dt'} |c_{n,k}|^{2} dt' \nonumber \\
&= E_{g} \left[ \int_{0}^{t}  \rho_{0}(t') \frac{d} {dt'} |\braket{g|k_{0}(t')}|^{2} dt' + \int_{0}^{t}  \rho_{1}(t') \frac{d} {dt'} |\braket{g|k_{1}(t')}|^{2} dt' \right] \nonumber \\ &+E_{e} \left[ \int_{0}^{t} \rho_{0}(t') \frac{d} {dt'} |\braket{e|k_{0}(t')}|^{2} dt' + \int_{0}^{t}  \rho_{1}(t') \frac{d} {dt'} |\braket{e|k_{1}(t')}|^{2} dt' \right] \nonumber \\
&= \frac{1}{4} (E_{e}-E_{g}) \left[\frac{1}{2} \log \left(-e^t+e^{2 t}+1\right)-\sqrt{3} \tan
   ^{-1}\left(\frac{2 e^t-1}{\sqrt{3}}\right)- t +\frac{\pi }{2 \sqrt{3}}\right].
\end{align}
This result is also plotted in Fig. 2 along with that for the internal energy, $\Delta U (t) = \mathcal{Q}(t) + \mathcal{C}(t)$.


\begin{thebibliography}{7}
\bibitem{SM} Further details are given in the Appendixes.
\bibitem{comment} In this specific quantum dynamics, the state of the system is initially pure, becomes partially mixed, and ends up pure again.
\end{thebibliography}
\end{document}